\documentclass[10pt]{iopart}
\pdfoutput=1

\usepackage{adjustbox}
\usepackage{algorithm}
\usepackage{algorithmic}
\usepackage[backend=biber, style=ieee]{biblatex}
\usepackage{booktabs}
\usepackage{etoolbox}
\usepackage{graphicx}
\usepackage{hyperref}
\usepackage{iopams}
\usepackage[switch]{lineno}
\usepackage{multirow}
\usepackage{silence}
\usepackage{siunitx}
\usepackage{subcaption}
\usepackage{tabularx}
\usepackage{tikz}
\usepackage{xcolor}

\usepackage{todonotes}

\usepackage[capitalize,noabbrev]{cleveref}

\graphicspath{{figures/}}

\definecolor{hgf-blue}{RGB}{0,90,160}
\definecolor{hgf-green}{RGB}{140,180,35}
\definecolor{hgf-gray}{RGB}{90,105,110}
\definecolor{hgf-purple}{RGB}{160,35,90}
\definecolor{hgf-red}{RGB}{210,50,100}
\definecolor{hgf-orange}{RGB}{240,120,30}
\definecolor{hgf-yellow}{RGB}{255,210,40}
\definecolor{hgf-turqoise}{RGB}{80,200,170}
\definecolor{hgf-dark-green}{RGB}{50,100,105}

\newtheorem{theorem}{Theorem}
\newtheorem{lemma}[theorem]{Lemma}

\makeatletter
\providerobustcmd*{\bigcupdot}{%
    \mathop{%
        \mathpalette\bigop@dot\bigcup
    }%
}
\newrobustcmd*{\bigop@dot}[2]{%
    \setbox0=\hbox{$\m@th#1#2$}%
    \vbox{%
        \lineskiplimit=\maxdimen
        \lineskip=-0.7\dimexpr\ht0+\dp0\relax
        \ialign{%
            \hfil##\hfil\cr
            $\m@th\cdot$\cr
            \box0\cr
        }%
    }%
}
\makeatother

\begin{document}

\title[Learning Tree Structures from Leaves For Particle Decay Reconstruction]{Learning Tree Structures from Leaves For Particle Decay Reconstruction}

\author{ %
    James Kahn\textsuperscript{1,2}, Ilias Tsaklidis\textsuperscript{3}, Oskar Taubert\textsuperscript{1,2}, Lea Reuter\textsuperscript{4}, Giulio Dujany\textsuperscript{5}, Tobias Boeckh\textsuperscript{3}, Arthur Thaller\textsuperscript{6}, Pablo Goldenzweig\textsuperscript{4}, Florian Bernlochner\textsuperscript{3}, Achim Streit\textsuperscript{2}, Markus G\"{o}tz\textsuperscript{1,2}
}

\address{ %
    \textsuperscript{1}Helmholtz AI\\
    \textsuperscript{2}Steinbuch Centre for Computing (SCC), Karlsruhe Institute of Technology (KIT), Germany\\
    \textsuperscript{3}Physikalisches Institut der Rheinischen Friedrich-Wilhelms-Universität Bonn, Germany\\
    \textsuperscript{4}Insitute for Experimental Particle Physics (ETP), Karlsruhe Institute of Technology (KIT), Germany\\
    \textsuperscript{5}Universit\'e de Strasbourg, CNRS, IPHC, UMR 7178, 67037 Strasbourg, France\\
    \textsuperscript{6}Aix Marseille Universit\'e, CNRS, IN2P3, CPPM, 13288 Marseille, France
}
\ead{james.kahn@kit.edu}
\vspace{10pt}
\begin{indented}
\item[]August 2022
\end{indented}

\begin{abstract}
In this work, we present a neural approach to reconstructing rooted tree graphs describing hierarchical interactions, using a novel representation we term the Lowest Common Ancestor Generations (LCAG) matrix.
This compact formulation is equivalent to the adjacency matrix, but enables learning a tree's structure from its leaves alone without the prior assumptions required if using the adjacency matrix directly.
Employing the LCAG therefore enables the first end-to-end trainable solution which learns the hierarchical structure of varying tree sizes directly, using only the terminal tree leaves to do so.
In the case of high-energy particle physics, a particle decay forms a hierarchical tree structure of which only the final products can be observed experimentally, and the large combinatorial space of possible trees makes an analytic solution intractable.
We demonstrate the use of the LCAG as a target in the task of predicting simulated particle physics decay structures using both a Transformer encoder and a Neural Relational Inference encoder Graph Neural Network.
With this approach, we are able to correctly predict the LCAG purely from leaf features for a maximum tree-depth of $8$ in $92.5\%$ of cases for trees up to $6$ leaves (including) and $59.7\%$ for trees up to $10$ in our simulated dataset.
\end{abstract}

%
\vspace{2pc}
\ioptwocol

\section{Introduction}
\label{sec:introduction}

Can the structure of a tree graph be determined from its leaves alone?
If all the information necessary to construct each parent node is contained in its children, then intuitively the answer is yes.
This is often the scenario we are presented with in real world situations, where a tree represents the chronological interactions of objects.
Such structures arise, for example, in semantic tree parsing, where the logic of a sequence like a sentence or mathematical equation is represented as a semantic tree, citation relevance hierarchies, where related publications are represented by a citation connection tree, or the decay of sub-atomic particles, where the topology of the decay process is determined by physical laws.
In these situations, we are limited to observing the properties of the leaves of such trees and inferring details about its structure from the leaves alone.

Well understood composition laws can allow us to reconstruct tree relations by attempting all reasonable combinations of children and verifying the physical validity of parents formed.
In situations involving a large number of tree topologies, however, the combinatorial complexity of such solutions makes them infeasible.
The ability to efficiently learn the relational structure of a tree from its leaves alone provides a means of reducing this complexity to a tractable level, where domain-specific knowledge can then be applied to fill in the constituent properties.

Applying deep learning to the task of tree reconstruction requires an effective representation of the structural information.
This representation is essential to providing a tangible learning target.
Furthermore, we are often presented with scenarios in which there is no prior knowledge about the depth of the tree or the degrees of nodes within.
For this, we propose the Lowest Common Ancestor Generation (LCAG) matrix, a novel tree representation which encodes a tree's structure in a compact representation suitable for use as a training target.
We demonstrate the use of the LCAG in learning how to reconstruct tree structures purely from leaves 
in one of the aforementioned examples: sub-atomic particle decays.

\begin{figure}[ht]
    \centering
    \includegraphics[width=0.9\columnwidth]{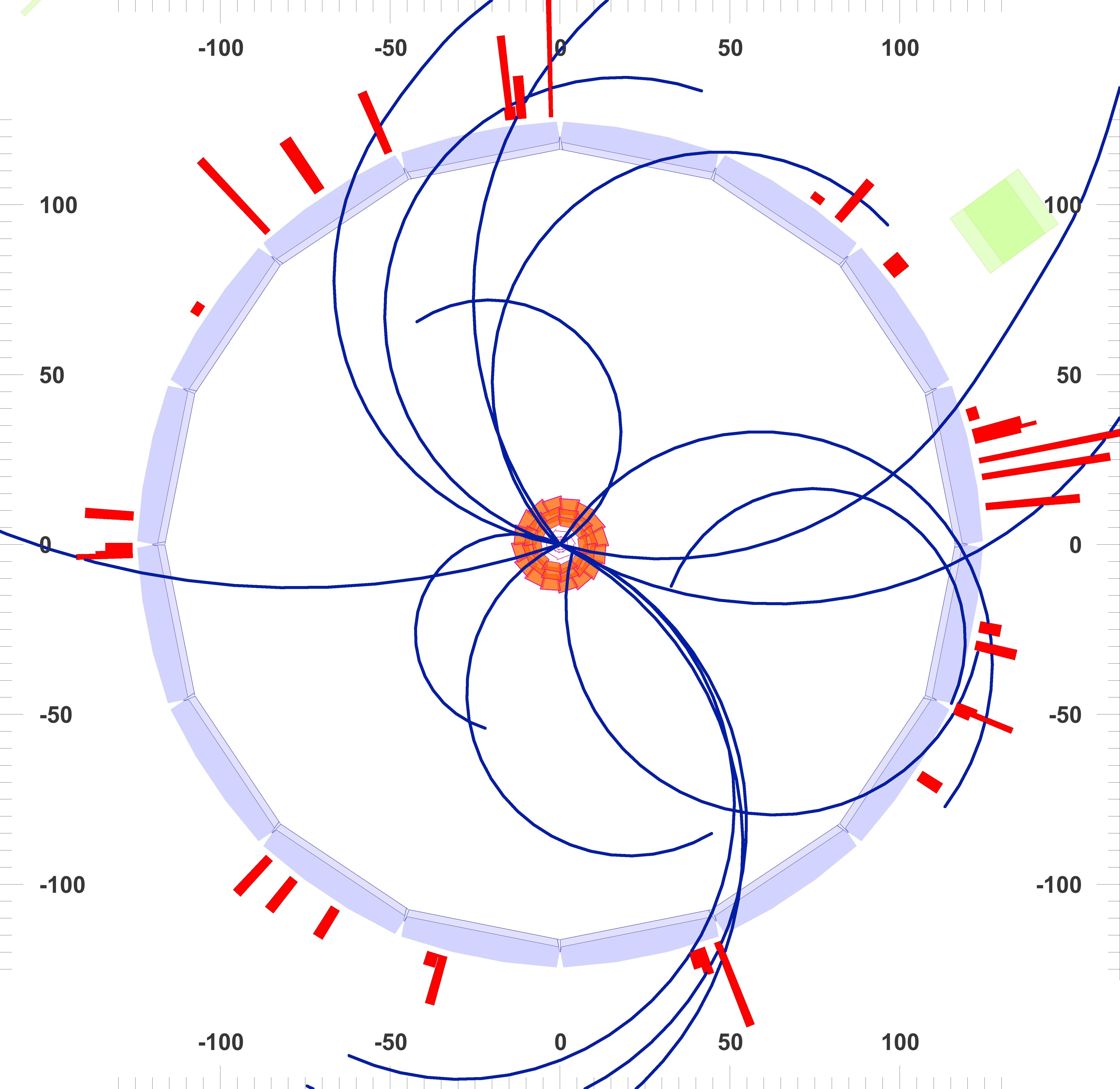}
    \caption{Example cross-sectional view (beam direction of the detector not shown) of a particle decay from the Belle~II experiment (image created using~\cite{collaborationBelleIIAnalysis2021}).
    Blue lines show the path of charged particles through the detector.
    The red and blue rings are detector components, and red and green towers represent energy deposits from detected particles.
    }
    \label{fig:event-display-2D}
\end{figure}

Particle colliders, such as the Large Hadron Collider (LHC)~\cite{evansLHCMachine2008} and the Belle~II experiment~\cite{abe2010belle}, study the fundamental laws of nature by accelerating and colliding particles at close to the speed of light.
The collisions that occur produce heavy subatomic particles which rapidly decay into lighter particles travelling away from the collision point. 
Subsequent decays may follow, until lighter, stable particles live long enough to reach a detection device.
In \cref{fig:event-display-2D}, an example of the trajectories of charged particles originating from a collision in the Belle~II experiment is shown.
This chronological structure can be naturally represented as a graph, or more specifically, as a rooted tree.
The terminal leaves of the tree consist of the particles that reach the detector equipment, the intermediate nodes their ancestors, and the edges of the graph their parent-child relations.
The topology and constituents of the tree are dictated by physical laws, for example conservation of energy and momentum.
In order to learn the interactions of this particle decay tree, the employed model must infer these physical laws, and hence predict the structure of the tree.

The contributions of this work are:
\begin{itemize}
    \item
    We introduce the Lowest Common Ancestor Generation (LCAG) matrix, a compact representation of rooted tree graph structures suitable for machine learning-based reconstruction of tree structures from leaves.
    \item
    We present a modification of neural relational inference for graph learning, tailored to predicting entire graph structures in a supervised setting.
    \item
    We use the LCAG as a training target for the neural reconstruction of simulated particle physics decays.
    In doing so, we demonstrate how the LCAG enables the first end-to-end trainable solution which learns the hierarchical structure of varying tree sizes directly, using only the terminal tree leaves to do so.
    \item
    We open-source the PyTorch implementation for this work, experiments, and trained models used\footnote{\url{https://github.com/Helmholtz-AI-Energy/BaumBauen}}, as well as the data~\cite{kahnLowestCommonAncestor2022a}.
\end{itemize}

\section{Related Work}
\label{sec:related-work}

Learning and exploiting the hierarchy of graphs has drawn attention in a variety of machine learning domains in recent years.
Works in the field of neural machine translation and semantic parsing~\cite{liGraphtoTreeNeuralNetworks2020a,miuraIntegratingTreeStructures2018}, or citation count prediction and review models~\cite{qiaoTreeStructureAwareGraph2020a}, learn the hierarchy of rooted tree structures.
Graph pooling~\cite{ying2018hierarchical} was introduced to create hierarchical representations of Graph Neural Networks~(GNNs) themselves to boost performance on graph classification tasks.
However, these all address the problem of predicting hierarchical structures of known and fixed depth.
In \cite{zhong2020hierarchical}, the authors utilized message passing functions to create hierarchies within a graph for community detection, but the case of a predefined tree structure remains.

The field of relational inference concerns the task of learning the interaction dynamics within a system from observational data.
Work in the field includes the Neural Relational Inference (NRI) model~\cite{kipf2018neural}, and the Hierarchical Relational Inference (HRI)~\cite{stanic2020hierarchical}.
Both works adopt an unsupervised approach to infer the relational structure of graphs, with the former operating on graphs representing dynamical systems, and the latter extending it to images of the systems. 
Other supervised approaches take the form of edge labelling tasks or edge weight prediction, and aim at binary edge classification~\cite{kim2019edgelabeling}, node clustering~\cite{finley2005supervised,wang2020unifying}, or graph classification~\cite{ranjanASAPAdaptiveStructure2020a}.
These supervised methods mainly focus on pairs of edges without dealing with global hierarchies inside the graphs.

For the special case that a graph is fully connected, i.e., all nodes connected to all others, an approach that relates all inputs to all others is appropriate.
The self-attention mechanism employed by the Transformer neural network~\cite{vaswani2017attention} operates as a GNN for such a case, modelling the pair-wise interactions between inputs.

In the field of particle physics, machine learning has begun to play a prominent role~\cite{guestDeepLearningIts2018,larkoskiJetSubstructureLarge2020,albertssonMachineLearningHigh2018}, and more recently GNNs have received a lot of attention~\cite{shlomi2021graph,duarteGraphNeuralNetworks2020}. 

GNNs have been investigated for improving direct particle reconstruction methods\footnote{Particle reconstruction is the task of using raw information about the energy deposited in the detector devices to obtain the physical properties of a particle candidate.}. 
A GNN-based implementation of the Particle Flow algorithm~\cite{sirunyanParticleflowReconstructionGlobal2017,aaboudJetReconstructionPerformance2017}, a key piece of software at the LHC which combines information from multiple detector hardware components to identify individual particles, is now being explored as an end-to-end trainable alternative to the existing rule-based approach~\cite{pataMLPFEfficientMachinelearned2021,mokhtarExplainingMachinelearnedParticleflow2021}.
The HEP.TrkX~\cite{farrellHEPTrkXProject2017}, and subsequent Exa.TrkX~\cite{juPerformanceGeometricDeep2021} have demonstrated the power of GNNs in the task of tracking individual charged particles at the LHC.
\cite{juGraphNeuralNetworks2020} further reinforced the ability of GNNs to learn the kinematics of particles based on detected information alone.
These works focus on individual particle tracks through detector equipment only, and do not deal with the decay tree structures of the particles themselves.

Jet vertex finding has been of particular interest to the LHC experiments and involves the labeling of detected particles according to their vertex in hadronic particle decays (jets).
\cite{henrion2017neural} used message-passing neural networks to learn the adjacency matrices for jet decays directly.
\cite{shlomi2021secondary} applied a range of Graph Neural Networks to what was effectively an edge classification task, and used the predicted binary edge labels to identify siblings in the jet decay tree structure.
However, these approaches presuppose the graph structure of the jets being reconstructed, for example as containing a primary and multiple secondary vertices.
JEDI-net~\cite{morenoJEDInetJetIdentification2020} attempts to identify the topology of a given jet, but only goes as far as classifying which topology, and not assigning specific particles vertices within.
ParticleNet~\cite{quJetTaggingParticle2020} applies a Dynamic Graph Convolutional Neural Network~\cite{wangDynamicGraphCNN2019} to identify the root particle of jets, treating the detected particles as a point cloud.
The transformer-based ParT model~\cite{quParticleTransformerJet2022a} explored using the set of individual particle kinematic and interaction information directly for the same task.
The above works recognize the importance of a meaningful input data representation, namely, treating detected particles as an unordered set, but fail to leverage a similarly meaningful representation for the labels as well. 
However, they all again reinforce that GNNs are an appropriate choice for learning the complex dynamics of particle decays.

The Full Event Interpretation (FEI)~\cite{keck2019full} software used by the Belle~II collaboration attempts to reconstruct entire particle decay trees using a sequence of individually trained Boosted Decision Trees (BDTs).
However, the specific decay processes targeted by the BDTs are hard-coded.
This both limits the total possible decay processes that can be reconstructed by the FEI, and requires domain-driven optimisation in every step of the tree reconstruction.
Furthermore, as the BDTs are trained independently from one another, this approach is prone to error accumulation, unlike a single, end-to-end trainable solution.

Our work attempts to resolve the constraints of the aforementioned works by introducing a new, compact representation of hierarchical trees that can be used as a single target label.
The goal being to no longer require assumptions about the structure, i.e., the depth or number of vertices, and leverage the power of a fully end-to-end trainable approach.
We do so in the context of particle decays, using GNNs inspired by previous works.

\section{Learning Tree Structures from Leaves}
\label{sec:learning-tree-struct}

A graph is a data structure that can be used to model a set of interdependent entities (nodes) and their relationships (edges).
More formally, adopting the notation of \cite{wu2021survey}, given a set of objects and their relations, a graph $\mathcal{G}$($\mathcal{V}$, $\mathcal{E}$), where $\mathcal{V}$ is the set of $\upsilon_{i}$ nodes and $\mathcal{E}$ the set of $e_{ij} = (\upsilon_{i}, \upsilon_{j})$ edges, expresses the relational structure of the system.
The adjacency matrix \textbf{A} of a graph of $n$ nodes is an $n \times n$ binary matrix, with $A_{ij} = 1$ if $e_{ij} \in \mathcal{E}$ and $A_{ij} = 0$ if $e_{ij} \notin \mathcal{E}$.
Nodes are described by the feature vector $\mathbf{x}_i \in \mathbf{R}^{d}$, given $d$ features of node $\upsilon_{i}$, and edges by the feature vector $\mathbf{x}_{(i,j)} \in \mathbf{R}^{c}$ for $c$ edge features between nodes $\upsilon_{i}$ and $\upsilon_{j}$.
Rooted trees are a special case of graphs that are acyclic and have one node designated as the root.

We can therefore abstract the hierarchical structure problem described in \cref{sec:introduction} using rooted trees as follows:
given a set of $k$ leaf node, represented by a corresponding set of vertices $\{ v_i \vert 1 \leq i \leq k\}$ and features $\{ x_i \}$, find the missing intermediate nodes and edges in the form of an adjacency matrix $A$.
If we are to learn $A$, we must use a solution representation that satisfies the following:
first, since there is in general no implicit order to the leaf nodes in graph-based problems, the solution representation must be permutation invariant (or at least permutation aware);
second, as the total number of nodes in a given tree is initially unknown, we also require a data representation for the solution whose size depends only on the number of leaves.
The adjacency matrix itself only satisfies the first condition, but requires the total number of nodes in $\mathcal{G}$ to be known a priori in order to define the correct number of rows and columns.
In the following section we present a compact representation, equivalent to the adjacency matrix for the case of hierarchical trees, that satisfies these requirements with minimal constraints.

In this work we limit ourselves strictly to learning the hierarchical tree structure.
We make the explicit assumption that once the structure is predicted, either non-leaf-node labels can be inferred in a later step or are not of interest.
Our goal is to provide a solution for cases in which the combinatorial explosion in finding the tree structure is a major limiting factor preventing simply attempting all feasible combinations.

\subsection{Lowest Common Ancestor Generation (LCAG) Tree Representation}
\label{sec:lca-tree-rep}

To satisfy the problem requirements outlined above, we employ a modified representation of the lowest common ancestor (LCA) matrix representation of rooted trees~\cite{ahoFindingLowestCommon1973}.
Given two nodes in a graph, $a$ and $b$, the lowest common ancestor is defined as the deepest node that is an ancestor of both $a$ and $b$, deepest in this case meaning farthest graph distance, $\textrm{dist}$, from the root node $r$, i.e.:

\begin{eqnarray*}
    \textrm{LCA}(a,b)=\textrm{argmax}_{p} \{\textrm{dist}(p,r) |\, &p\in\textrm{path}(a,r)\land \\
    &p\in \textrm{path}(b,r)\}\ .
\end{eqnarray*}

\begin{figure}
    \centering
    \adjustbox{width=\linewidth}{%
        \begin{tikzpicture}[font=\sffamily\small]
    \def\cs{0.7}
    \def\sw{0.8}
    \def\sh{2.0}
    \def\grid{0.4}

    
    
    
    \node[circle, draw=hgf-green, fill=hgf-green!20!white, minimum size=\cs cm, inner sep=0] (Anode) at (0.0, 0.0) {A};
    
    \node[circle, draw=hgf-red, fill=hgf-red!20!white, minimum size=\cs cm, inner sep=0] (Bnode) at (-1.5 * \sw, -1 * \sh) {B};
    \node[circle, draw=hgf-red, fill=hgf-red!20!white, minimum size=\cs cm, inner sep=0] (Cnode) at (1.5 * \sw, -1 * \sh) {C};
    
    \node[circle, draw=hgf-blue, fill=hgf-blue!20!white, minimum size=\cs cm, inner sep=0] (vnode) at (-2.0 * \sw, -2 * \sh) {v};
    \node[circle, draw=hgf-blue, fill=hgf-blue!20!white, minimum size=\cs cm, inner sep=0] (wnode) at (-1.0 * \sw, -2 * \sh) {w};
    \node[circle, draw=hgf-blue, fill=hgf-blue!20!white, minimum size=\cs cm, inner sep=0] (xnode) at (0.0 * \sw, -2 * \sh) {x};
    \node[circle, draw=hgf-blue, fill=hgf-blue!20!white, minimum size=\cs cm, inner sep=0] (ynode) at (1.0 * \sw, -2 * \sh) {y};
    \node[circle, draw=hgf-blue, fill=hgf-blue!20!white, minimum size=\cs cm, inner sep=0] (znode) at (2.0 * \sw, -2 * \sh) {z};
    
    \draw[-latex] (Anode) -- (Bnode);
    \draw[-latex] (Anode) -- (Cnode);
    \draw[-latex] (Anode) -- (xnode);
    \draw[-latex] (Bnode) -- (vnode);
    \draw[-latex] (Bnode) -- (wnode);
    \draw[-latex] (Cnode) -- (ynode);
    \draw[-latex] (Cnode) -- (znode);
    
    \def\colors{{v}/hgf-blue!20!white, {B}/hgf-red!20!white, {A}/hgf-green!20!white, {A}/hgf-green!20!white, {A}/hgf-green!20!white, {B}/hgf-red!20!white, {w}/hgf-blue!20!white, {A}/hgf-green!20!white, {A}/hgf-green!20!white, {A}/hgf-green!20!white,    {A}/hgf-green!20!white, {A}/hgf-green!20!white, {x}/hgf-blue!20!white, {A}/hgf-green!20!white, {A}/hgf-green!20!white, {A}/hgf-green!20!white, {A}/hgf-green!20!white, {A}/hgf-green!20!white, {y}/hgf-blue!20!white, {C}/hgf-red!20!white, {A}/hgf-green!20!white, {A}/hgf-green!20!white, {A}/hgf-green!20!white, {C}/hgf-red!20!white, {z}/hgf-blue!20!white}
    
    \foreach[count=\i] \t/\c in \colors {
        \node[minimum height=\grid cm + 0.02 cm, minimum width=\grid cm + 0.02, fill=\c, inner sep=0] at ({Mod(\i - 1, 5.0) * \grid + 2.4 + 1.5 * \grid}, {-floor((\i - 1)/ 5.0) * \grid + 0.25 * \cs + 0.03}) {\scriptsize\t};
    }
    
    \def\colors{0/hgf-blue!20!white, 1/hgf-red!20!white, 2/hgf-green!20!white, 2/hgf-green!20!white, 2/hgf-green!20!white, 1/hgf-red!20!white, 0/hgf-blue!20!white, 2/hgf-green!20!white, 2/hgf-green!20!white, 2/hgf-green!20!white, 2/hgf-green!20!white, 2/hgf-green!20!white, 0/hgf-blue!20!white, 2/hgf-green!20!white, 2/hgf-green!20!white, 2/hgf-green!20!white, 2/hgf-green!20!white, 2/hgf-green!20!white, 0/hgf-blue!20!white, 1/hgf-red!20!white, 2/hgf-green!20!white, 2/hgf-green!20!white, 2/hgf-green!20!white, 1/hgf-red!20!white, 0/hgf-blue!20!white}
    
    \foreach[count=\i] \t/\c in \colors {
        \node[minimum height=\grid cm + 0.02 cm, minimum width=\grid cm + 0.02, fill=\c, inner sep=0] at ({Mod(\i - 1, 5.0) * \grid + 2.4 + 1.5 * \grid}, {-floor((\i - 1)/ 5.0) * \grid + 0.25 * \cs + 0.03 - 2.4}) {\scriptsize\t};
    }
    
    \def\names{v, w, x, y, z}
    \foreach[count=\i] \n in \names {
        \node at (2.4 + \grid / 2 + \i * \grid, - 4.2) {\scriptsize\n};
        
        \node[align=left] at (2.5, 0.0 + \cs / 2 + 0.75 * \grid - \i * \grid) {\scriptsize\n};
        
        \node[align=left] at (2.5, -2.1 + \cs / 2 - \i * \grid) {\scriptsize\n};
    }
    
    \draw[step=\grid, black] (2.8 - 0.01, \cs / 2 + 0.05) grid (5 * \grid + 2.8, -4 * \grid);
    \draw[step=\grid, black] (2.8 - 0.01, -2.4 + \cs / 2 + 0.05) grid (5 * \grid + 2.8, -4 * \grid - 2.4);
    
    \draw[densely dashed, hgf-gray!50!white] (-3.8, -0.5 * \sh) -- (2.0, -0.5 * \sh);
    \draw[densely dashed, hgf-gray!50!white] (-3.8, -1.5 * \sh) -- (2.0, -1.5 * \sh);
    
    \node[align=left, anchor=north west] at (-3.8, 0.25 * \sh) {\bfseries\scriptsize Root\\\scriptsize 2$^{nd}$ gen.};
    
    \node[align=left, anchor=north west] at (-3.8, -0.55 * \sh) {\bfseries\scriptsize Decay products\\\bfseries\scriptsize (unknown)\\\scriptsize 1$^{st}$ gen.};
    
    \node[align=left, anchor=north west] at (-3.8, -1.55 * \sh) {\bfseries\scriptsize Stable particles\\\bfseries\scriptsize (leaves)\\\scriptsize 0$^{th}$ gen.};
\end{tikzpicture}
    }
    \caption{
    Example of a hierarchical tree representing a particle decay (left), its corresponding LCA (top right), and LCAG matrix (bottom right).
    Colours are added for visual clarity only, and node labels are simply examples to uniquely identify nodes and do not correspond to real particle names.
    }
    \label{fig:example-event}
\end{figure}
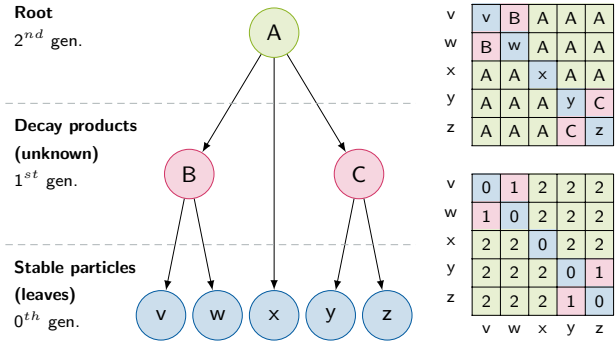

\Cref{fig:example-event} shows an example tree representing an arbitrary particle decay process (left) and its corresponding LCA matrix (top right), with generations up from the leaves labelled on the far left.
Letters indicate arbitrary node labels, and the colours are simply added for clarity and do not hold further meaning\footnote{We note here that the node labels in an LCA are unique identifiers of each node, and do not correspond, in the physics example, to the particle types.}.
In order for this representation to describe a graph unambiguously, it must be a tree where any non-leaf node has \emph{two or more} children.
This way the set of nodes that are elements of the LCA matrix is identical to $\mathcal{V}$.
This assumption bounds the maximum depth of the tree to $\mathcal{O}(\log_2 k)$ for $k$ leaves.
Note that each entry in the LCA matrix encodes information about an unordered pair of vertices, and as such is symmetric.

In order to express the lowest common ancestor matrix in a form appropriate for machine learning, we propose the lowest common ancestor generation (LCAG) matrix, in which each entry of the LCA matrix is replaced with its corresponding generation in the tree.
\Cref{fig:example-event} (bottom right) shows the LCAG corresponding to the example tree.
We see here a clear distinction between the LCA and LCAG: while the LCA requires unique node labels to identify ancestors, the LCAG is a relaxed form which allows the same identifier (generation) to represent multiple ancestors.
In other words, the LCA is surjective to the LCAG.
The underlying assumption for making this modification is that if there exist structural rules which allow the inference of the tree structure from the leaf nodes alone, then these rules can also be used on the inferred structure to deduce unseen node labels and hence recover the LCA.
Relating this to the example particle decay process in \cref{fig:example-event}, if we know that two particles $\textrm{v}$ and $\textrm{w}$ come from the same parent, summing these energies would tell us the mass, and hence the node label of the parent $\textrm{B}$.

For the LCAG matrix, we adopt a pull-down convention:
every node in the tree resides at the lowest possible generation, with all leaves fixed at the zeroth generation.
Formally:

\begin{eqnarray*}
    \textrm{LCAG}_{ij}(\mathcal{G}) &= g(\textrm{LCA}_{ij}(\mathcal {G}))\ , \\
    g(u) &=\left\{\begin{array}{l@{\quad}l} 
        0, & \textrm{if $u$ is a leaf} \\
        \max\limits_{v \in \mathcal{C}(u)} \{g(v)\} + 1, & \textrm{otherwise}
    \end{array}\right.
\end{eqnarray*}
where $\mathcal{C}(u)$ is the set of children of node $u$.

\begin{lemma}
\label{lemma:lcag_lemma}
The LCAG matrix uniquely describes any rooted tree isomorphism class for trees in which every non-leaf vertex has two or more children.
I. e.: If two trees $G = (V_G, E_G)$ and $H = (V_H, E_H)$ are described by the LCAG $L$, then $G$ and $H$ are isomorphic ($G \simeq H$).
\end{lemma}

We prove this by complete induction.
\cref{lemma:lcag_lemma} is trivially true for trees consisting of a single root node, described by the LCAG $\left[ 0\right]$.
We will now show that it also holds for any tree of height $m+1$ if it holds for trees of height $m$.
Consider two trees $G$ and $H$ of height $m+1$, that are described by the LCAG $L$.
Given any subtree $G_i$ of $G$ that includes all descendants of $\textrm{root}(G_i)$ in $V(G)$, $G_i$ is described by $L_{i}$ containing all entries of $L$ corresponding to the leaves of $G_i$.
The $L_{i}$ of all subtrees whose roots are siblings in $G$ and are direct descendants of $root(G)$ must be disjoint in $L$, since their LCA describes the parent of the roots of all $G_i$ in $G$.
The remaining entries of $L$ not in any $L_{i}$ are all by definition $m+1$.
The same set of $L_i$ correspond analogously to subtrees of $H$.
Therefore, for each $G_i$ we can find a corresponding subtree $H_i$ that is described the same $L_i$ in such a way that $\bigcupdot_i V(G_i) \cup \textrm{root}(G) = V(G)$ and $\bigcupdot_i V(H_i) \cup \textrm{root}(H) = V(H)$.
Because of \cref{lemma:lcag_lemma}, $G_i$ and $H_i$ are isomorphic, as their height is $m$ and there is an isomorphism $f_i: V(G_i) \rightarrow V(H_i)$.
We can find an isomorphism $f:V(G) \rightarrow V(H)$ by  $f(\textrm{root}(G)) = \textrm{root}(H)$ and $f(u) = f_{i}(u)$ for $u \in V(G_i)$.
Therefore, \cref{lemma:lcag_lemma} also holds for trees of height $m+1$.

\subsection{Converting LCAG from and to Adjacency Matrix}
\label{sec:converting-lca-adj}

In order to generate the adjacency matrix of a tree given its LCAG matrix, we proceed level by level, starting from the leaves.
We generate a list of leaves from the dimensions of the input LCAG.
We then compute the vertices in the next level by collecting all sets of siblings in the list, i.e., vertices whose lowest common ancestor is in the next level.
For each set of siblings, we remove them from the list and add their parent vertex to the list, keeping track of the lowest common ancestor of the newly added vertices and all other vertices in the list.
\Cref{alg:lcag-conversion} shows the pseudocode of the conversion algorithm.

To generate the LCAG from an adjacency matrix, we first find the set of leaves, and then fill the values of the matrix with the longest graph distance to the lowest common ancestor of the corresponding pair of leaf vertices.

\begin{algorithm}
    \caption{Pseudocode for the conversion of a LCAG gram matrices $L$ into a tree $T$, error handling for ill-formatted matrices omitted. Indices and ranges are zero-indexed.}
    \label{alg:lcag-conversion}
    \footnotesize
    \begin{algorithmic}
        \STATE {\bfseries Input:} Symmetric LCA(G) gram matrix $L$ ($n\times n$)
        \STATE {\bfseries Output:} Root of tree $T$ described by $L$
        \STATE $leaves \leftarrow [~]$
        \STATE $levels \leftarrow sorted(unique(L))$
        \STATE $total\_nodes \leftarrow n$
        \SHORTFORALL{$n$}{$leaves.append(Node(level=0))$}
            
        \COMMENT{level by level}
        \FOR{$level$ {\bfseries in} $[0,\ ...,\ levels.length]$}
            \FOR{$column$ {\bfseries in} $[0,\ ...,\ n]$}
                \FOR{$row$ {\bfseries in} $[column+1,\ ...,\ n]$}
                    \SHORTIF{$L[row, column]\neq level$}{continue}
                    \STATE $node\_a \leftarrow leaves[row]$
                    \STATE $node\_b \leftarrow leaves[column]$
                    \STATE $root\_a \leftarrow get\_root(node\_a)$
                    \STATE $root\_b \leftarrow get\_root(node\_b)$
                    
                    \COMMENT{same level, new node}
                    \IF{$root\_a.level< level+1$ \AND $root\_b.level< level+1$}
                        \STATE $parent \leftarrow Node(level=level+1)$\;
                        \STATE $parent.children.append([node\_a, node\_b])$\;
                        \STATE $total\_nodes \leftarrow total\_nodes +1$\;
                    \COMMENT{root\_b is older}
                    \ELSIF{$root\_a.level < level + 1$ \AND $root\_b == level + 1$}
                        \STATE $root\_a.children.append(root\_b)$
                    \COMMENT{vice versa}
                    \ELSE
                        \STATE $root\_b.children.append(root\_a)$
                    \ENDIF
                \ENDFOR
            \ENDFOR
        \ENDFOR
        \STATE $root\gets get\_root(leaves[0])$\;
    \end{algorithmic}
\end{algorithm}

\section{Experimental Setup and Results}

In this work, we demonstrate the method of learning the LCAG as a training target using a Graph Neural Network to predict a collection of simulated particle decays.
To the best of our knowledge this is the first end-to-end trainable solution which learns the hierarchical structure of varying tree sizes directly, using only the terminal tree leaves to do so.

\subsection{Data}
\label{sec:data}

Both the LHC and Belle~II simulated collision data are restricted to internal member access only, and existing, publicly accessible benchmarks such as the TrackML challenge~\cite{amroucheTrackingMachineLearning2020} and the PD4ML~\cite{benatoSharedDataAlgorithms2021} focus only on identifying individual particles or whole-graph classification tasks, not any form of decay tree reconstruction.
Therefore, we simulate our own experimental dataset using the Phasespace library~\cite{navarro2019phasespace}.
This library takes as input a tree structure in the form of parent-daughter relations, as well as the mass of each participating particle (node).
Decays are then simulated according to Monte Carlo phase space~\cite{james1968monte}, which obey the usual physical laws (conservation of energy, momentum, etc.).
The output is the four-momentum (energy and 3D momentum) of each participating particle, of which we use the leaf nodes for input to the network.
In a high-energy particle physics experiment, these leaf nodes correspond to particles detected by the experimental hardware, which a physicist would then use to reconstruct the originating decay.
We refer to different tree structures as topologies, and we refer to individual simulated trees (of any topology) as samples. 

The simulated datasets used by particle physics experiments for building similar types of reconstruction algorithms include only known topologies. 
For the type of usage of such algorithms, it is intentional that only those topologies that are well understood are reconstructed to avoid discrepancies between simulation (i.e. training) and measured data (i.e. inference). 
Therefore, it is reasonable to expect all possible topologies intended for reconstruction be present in the training dataset, with no generalisation to new, unseen topologies.

We simulate a synthetic particle decay dataset for our experiments, consisting of topologies with a common root particle of mass $100$ (arbitrary units).
Intermediate particles are selected at random with replacement from the following masses: $[90, 80, 70, 50, 25, 20, 10]$.
Final state particles, which make up the leaf nodes of generated topologies, are drawn with replacement from the following masses: $[1, 2, 3, 5, 12]$.
For each intermediate particle (including the root), we limit the minimum number of children to two, and the maximum five.
These masses and ranges of child particles are chosen to correspond to the relative mass scales and decay multiplicities observed in nature~\cite{zyla2020review}.
For particle physics decay processes seen at the Belle~II experiment, for example, the number of leaves is typically around ten or fewer, with existing decay reconstruction algorithms reflecting this\footnote{We discuss the case at the LHC, which produces significantly more leaves per collision event, later in \cref{sec:discussion}.}.

Tree topology creation is then as follows:
starting from the root particle a set of children are selected from the available intermediate and final state particles such that the sum of their masses totals less than the root, this process is then repeated for each child particle which is not a final state particle and so on until only final state particles remain.

The synthetic dataset consists of $200$ topologies in total, with $2000$ samples per topology for each of training, validation, and testing.
\Cref{fig:phasespace-results-perfectLCAG} (top) shows the distribution of simulated topologies, grouped according to the number of leaves and the total depth of the tree, which we use as a surrogate for topology complexity. 
All leaf node features are normalized to a normal distribution of mean zero, standard deviation one.
We do not enforce any ordering of the nodes and leave them unsorted as created in the dataset.

\subsection{Models}
\label{sec:models}

For the choice of neural network models, we require architectures that operate on unordered sets of vertices.
The architectures must be able to learn the inter-dependencies between vertices and model the underlying dynamics of the physical system.
We therefore select the encoder components of both the Transformer and Neural Relational Inference models.
All model implementations were made in \textit{PyTorch v1.8.1}~\cite{NEURIPS2019_9015} and trained using \textit{PyTorch-Ignite v0.4.4}~\cite{pytorch-ignite}.

\paragraph{Transformer} models have dominated natural language processing in recent years due to their ability to model complex, pair-wise interactions.
In the following experiments we employ the encoder component of the Transformer, followed by an outer-concatenation operation\footnote{An outer-concatenation involves duplicating the leaves along a new axis to produce a $(k \times d) \rightarrow (k \times k \times d)$ tensor and concatenating this with the transpose of itself.
This creates a $(k \times k \times 2d)$ dimension tensor of node pairs.} to project from node representations into pair-wise edge representations.
This is then fed through a fully-connected output head to produce the LCAG prediction.
We can interpret the LCAG as being a weighted adjacency matrix describing a graph of the leaves alone, where the integer weights represent the generation of the lowest common ancestor each pair of leaves.
Therefore we can apply the Transformer directly to the task of learning the LCAG.

\paragraph{NRI} was originally proposed by \cite{kipf2018neural} as an unsupervised Graph Neural Network in the form of a variational autoencoder.
In our experiments we use only the encoder component of the NRI, as we are performing supervised training.
The building blocks of an NRI layer consist of node-to-edge and edge-to-node message passing modules, followed by a fully-connected module each.
The latent representation then alternates between node-wise and edge-wise as it propagates through the network.
The encoder produces a discrete categorical distribution of the edge label predictions, which in our case are the entries of the LCAG. 
Each LCAG entry prediction is treated as an individual classification task with a softmax followed by an argmax function after the last layer.
The individual losses are averaged to produce the total loss.

\begin{figure*}[!ht]
    \centering
    \includegraphics[width=1.0\textwidth]{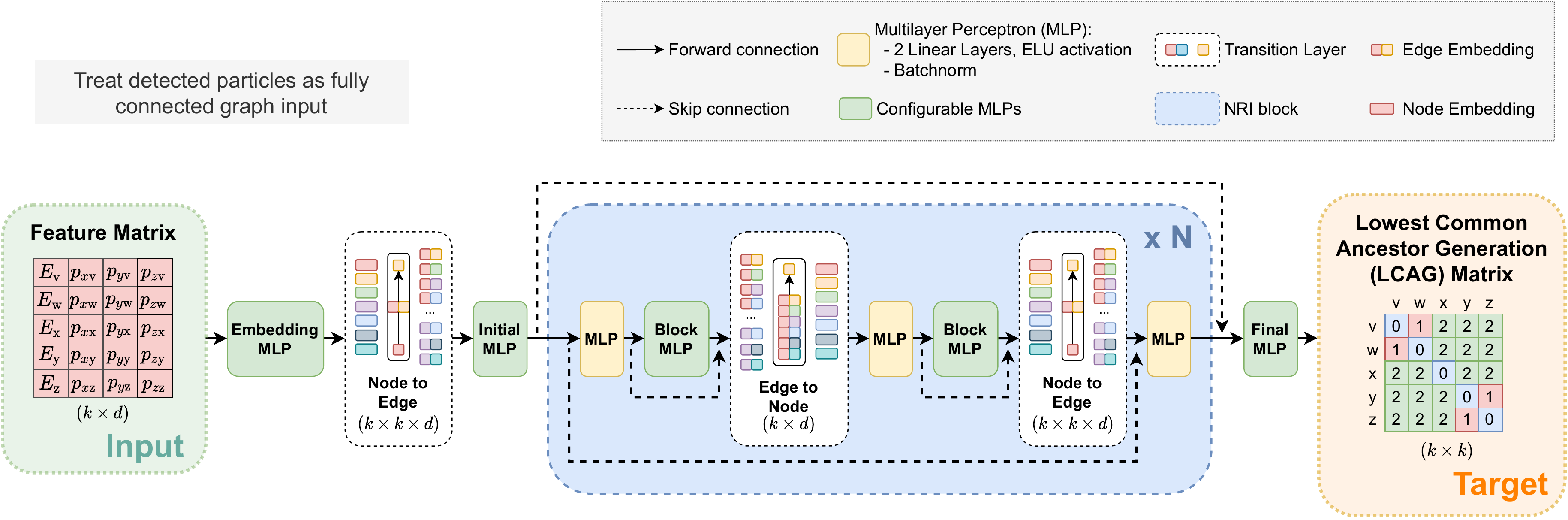}
    \caption{NRI encoder model architecture. The \emph{node-to-edge} and \emph{edge-to-node} layers shown are from \cite{kipf2018neural}, tensor shapes are shown only when dimension changes. The input feature matrix contains the four-momentum ($E, p_x, p_y, p_z$) of each leaf ($\textrm{v}, \textrm{w}, \ldots$).}
    \label{fig:NRI-architecture}
\end{figure*}

\Cref{fig:NRI-architecture} shows the complete NRI encoder architecture used in experiments.
To update the state embeddings between edge-to-node and node-to-edge transition layers, the NRI employs sequences of multilayer perceptrons (MLPs) containing two linear layers with ELU activations~\cite{clevertFastAccurateDeep2016} and batch normalisation~\cite{ioffeBatchNormalizationAccelerating2015}.
We extend the original NRI implementation to include a configurable number of additional MLPs both within the NRI blocks and at the beginning and end of the model to allow for added learning capacity.

The \emph{neural message-passing mechanism} in the NRI, used in the edge-to-node and node-to-edge layers, updates state embeddings by aggregating information across all connected edges in the graph (we adopt the notation of \cite{kipf2018neural}):

\begin{eqnarray*}
    \textrm{node-to-edge} &: \textbf{h}^{l}_{(i,j)} &= f_{e}^{l} \left( \left[ \textbf{h}_i^l, \textbf{h}_j^l \right] \right) \\
    \textrm{edge-to-node} &: \textbf{h}^{l+1}_i     &= f_{v}^{l} \left( \sum_{j} \textbf{h}_{(i,j)}^l  \right)
\end{eqnarray*}
where $\textbf{h}^l_i$ represents the state embedding of node $i$ at layer $l$, and $\textbf{h}^l_{(i,j)}$ the embedding of the edge connecting nodes $i$ and $j$.
$f_e$ and $f_v$ are feed-forward neural networks used to produce the edge and node embeddings.
In contrast to \cite{kipf2018neural}, we allow for self-interactions ($\textbf{h}_{(i,i)}^l$) in our experiments as we found this empirically to perform better.

\subsection{Evaluation Metrics}
\label{sec:evaluation-metrics}

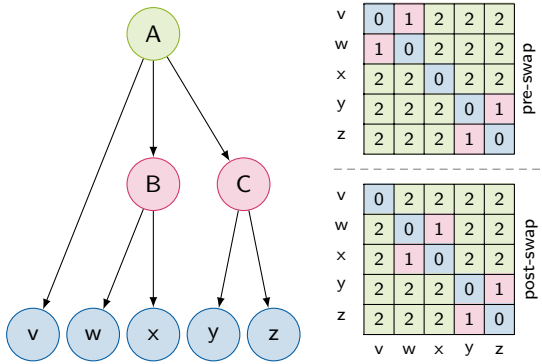
\begin{figure}[ht]
    \centering
    \begin{tikzpicture}[font=\sffamily\small]
    \def\cs{0.7}
    \def\sw{0.8}
    \def\sh{2.0}
    \def\grid{0.4}

    \node[circle, draw=hgf-green, fill=hgf-green!20!white, minimum size=\cs cm, inner sep=0] (Anode) at (0.0, 0.0) {A};
    
    \node[circle, draw=hgf-red, fill=hgf-red!20!white, minimum size=\cs cm, inner sep=0] (Bnode) at (0.0 * \sw, -1 * \sh) {B};
    \node[circle, draw=hgf-red, fill=hgf-red!20!white, minimum size=\cs cm, inner sep=0] (Cnode) at (1.5 * \sw, -1 * \sh) {C};
    
    \node[circle, draw=hgf-blue, fill=hgf-blue!20!white, minimum size=\cs cm, inner sep=0] (vnode) at (-2.0 * \sw, -2 * \sh) {v};
    \node[circle, draw=hgf-blue, fill=hgf-blue!20!white, minimum size=\cs cm, inner sep=0] (wnode) at (-1.0 * \sw, -2 * \sh) {w};
    \node[circle, draw=hgf-blue, fill=hgf-blue!20!white, minimum size=\cs cm, inner sep=0] (xnode) at (0.0 * \sw, -2 * \sh) {x};
    \node[circle, draw=hgf-blue, fill=hgf-blue!20!white, minimum size=\cs cm, inner sep=0] (ynode) at (1.0 * \sw, -2 * \sh) {y};
    \node[circle, draw=hgf-blue, fill=hgf-blue!20!white, minimum size=\cs cm, inner sep=0] (znode) at (2.0 * \sw, -2 * \sh) {z};
    
    \draw[-latex] (Anode) -- (Bnode);
    \draw[-latex] (Anode) -- (Cnode);
    \draw[-latex] (Anode) -- (vnode);
    \draw[-latex] (Bnode) -- (wnode);
    \draw[-latex] (Bnode) -- (xnode);
    \draw[-latex] (Cnode) -- (ynode);
    \draw[-latex] (Cnode) -- (znode);

    \def\colors{0/hgf-blue!20!white, 1/hgf-red!20!white, 2/hgf-green!20!white, 2/hgf-green!20!white, 2/hgf-green!20!white, 1/hgf-red!20!white, 0/hgf-blue!20!white, 2/hgf-green!20!white, 2/hgf-green!20!white, 2/hgf-green!20!white,    2/hgf-green!20!white, 2/hgf-green!20!white, 0/hgf-blue!20!white, 2/hgf-green!20!white, 2/hgf-green!20!white, 2/hgf-green!20!white, 2/hgf-green!20!white, 2/hgf-green!20!white, 0/hgf-blue!20!white, 1/hgf-red!20!white, 2/hgf-green!20!white, 2/hgf-green!20!white, 2/hgf-green!20!white, 1/hgf-red!20!white, 0/hgf-blue!20!white}

    \foreach[count=\i] \t/\c in \colors {
        \node[minimum height=\grid cm + 0.02 cm, minimum width=\grid cm + 0.02, fill=\c, inner sep=0] at ({Mod(\i - 1, 5.0) * \grid + 2.4 + 1.5 * \grid}, {-floor((\i - 1)/ 5.0) * \grid + 0.25 * \cs + 0.03}) {\scriptsize\t};
    }
    
    \def\colors{0/hgf-blue!20!white, 2/hgf-green!20!white, 2/hgf-green!20!white, 2/hgf-green!20!white, 2/hgf-green!20!white, 2/hgf-green!20!white, 0/hgf-blue!20!white, 1/hgf-red!20!white, 2/hgf-green!20!white, 2/hgf-green!20!white, 2/hgf-green!20!white, 1/hgf-red!20!white, 0/hgf-blue!20!white, 2/hgf-green!20!white, 2/hgf-green!20!white, 2/hgf-green!20!white, 2/hgf-green!20!white, 2/hgf-green!20!white, 0/hgf-blue!20!white, 1/hgf-red!20!white, 2/hgf-green!20!white, 2/hgf-green!20!white, 2/hgf-green!20!white, 1/hgf-red!20!white, 0/hgf-blue!20!white}
    
    \foreach[count=\i] \t/\c in \colors {
        \node[minimum height=\grid cm + 0.02 cm, minimum width=\grid cm + 0.02, fill=\c, inner sep=0] at ({Mod(\i - 1, 5.0) * \grid + 2.4 + 1.5 * \grid}, {-floor((\i - 1)/ 5.0) * \grid + 0.25 * \cs + 0.03 - 2.4}) {\scriptsize\t};
    }
    
    \def\names{v, w, x, y, z}
    \foreach[count=\i] \n in \names {
        \node at (2.4 + \grid / 2 + \i * \grid, - 4.2) {\scriptsize\n};
        
        \node[align=left] at (2.5, 0.0 + \cs / 2 + 0.75 * \grid - \i * \grid) {\scriptsize\n};
        
        \node[align=left] at (2.5, -2.1 + \cs / 2 - \i * \grid) {\scriptsize\n};
    }
    
    \draw[step=\grid, black] (2.8 - 0.01, \cs / 2 + 0.05) grid (5 * \grid + 2.8, -4 * \grid);
    \draw[step=\grid, black] (2.8 - 0.01, -2.4 + \cs / 2 + 0.05) grid (5 * \grid + 2.8, -4 * \grid - 2.4);
    
    \draw[gray, densely dashed] (2.4, -1.8) -- (7 * \grid + 2.4, -1.8);
    
    \node[anchor=north, rotate=90] at (6 * \grid + 2.4, -1.5 * \grid) {\scriptsize pre-swap};
    
    \node[anchor=north, rotate=90] at (6 * \grid + 2.4, -1.5 * \grid - 2.4) {\scriptsize post-swap};
    
    
    
    
\end{tikzpicture}
    \caption{Exchanging the parent of two leaves from the tree in \cref{fig:example-event} modifies two entries (in the upper/lower triangle) in the its LCAG matrix.}
    \label{fig:example-tree-swap-parent-lca}
\end{figure}

Similar to (non-ordinal) classification tasks, a predicted LCAG has no clear concept of an \emph{almost correct} tree, for example in the case when only one entry predictions is incorrect.
To see why this is the case, we can look at an example of what we would intuitively identify as two similar trees, i.e., almost isomorphic, and see that they are not related by valid LCAG representations.
Consider the simple operation of exchanging the parent of the nodes $\textrm{x}$ and $\textrm{v}$ in the example from \cref{fig:example-event}.
\Cref{fig:example-tree-swap-parent-lca} demonstrates how this alters two LCAG entries in the process (considering only the upper or lower triangle due to symmetry).
Changing only one of the LCAG entries would result in what we consider an \emph{invalid} LCAG: one that does not represent a coherent tree structure.
We therefore introduce the metric $\mathrm{perfectLCAG}$ to evaluate the rate of correct LCAG predictions, which is the ratio of correctly predicted LCAGs to the total number of samples.

\subsection{Experiments}
\label{sec:phasespace-exp}

We optimized and evaluated both the Transformer and the NRI encoder on the simulated dataset described above (cf. \cref{sec:data}).
To find the optimal hyperparameters for both models, we used the \textit{Optuna v2.10}~\cite{akibaOptunaNextgenerationHyperparameter2019} library with a Tree-structured Parzen Estimator~\cite{bergstraAlgorithmsHyperparameterOptimization2011} optimisation approach.
The objective value maximized is the $\mathrm{perfectLCAG}$ score on validation samples.
All trainings were performed on a single NVIDIA A100 GPU with $\SI{40}{GB}$ memory.

We performed an initial hyperparameter search on the parameters shown in \cref{tab:phasespace-hyperparams}.
Namely, we explored how the network performance changes when varying the building blocks, i.e., the attention layers in the Transformer or the blocks in the NRI, as well as the overall capacity of the models, i.e., the number of neurons in fully-connected layers, and additional initial/final MLPs.
As this is a multi-target classification task, i.e., one target per LCAG entry, where getting all LCAG entry predictions correct is more important than getting only a subset correct with a high confidence (as cross-entropy loss does) we also included focal loss~\cite{linFocalLossDense2017} in the search.

\begin{table}[htb]
    \caption{Phasespace hyperparameter search parameters. Asterisks indicate parameters used exclusively in the larger-dimension, follow-up search performed on the NRI.}
    \label{tab:phasespace-hyperparams}
        \centering
        \scriptsize
        \begin{tabularx}{\linewidth}{llX}
            \toprule
            \textbf{Model} & \textbf{Parameter}         & \textbf{Values} \\
            \midrule
            \multirow{6}{*}{NRI} & No. NRI blocks & [1, 2, 4, 6] \\
            & Block MLP layers & [0, 2, 4] \\
            & Feed-forward width & [128, 256, 512, 768*, 1024*, 2048*] \\
            & Initial MLP layers & [0*, 1, 2, 4] \\
            & Final MLP layers & [0*, 1, 2, 4] \\
            & Loss & [cross-entropy, focal] \\
            \midrule
            \multirow{5}{*}{Transformer} & No. attentions & [1, 2, 4, 6] \\
            & No. attention heads & [4, 8, 16] \\
            & Feed-forward width & [256, 512, 1024] \\
            & Final MLP layers & [1, 2, 4] \\
            & Loss & [cross-entropy, focal] \\
            \bottomrule
    \end{tabularx}
\end{table}

All hyperparameter search training were performed for a maximum of $100$ epochs, a batch size of $128$, dropout rate of $0.3$, random seed of $100$, and with class weights applied, using the Adam optimizer with a learning rate of $0.001$.
    
The results of the searches, shown for each individual hyperparameter, are shown in the supplementary material in \cref{fig:NRI-hyperparam,fig:transformer-hyperparam}.
Our experiments show the NRI to significantly outperform the Transformer encoder, reaching an average validation $\mathrm{perfectLCAG}$ of around $46\%$, compared to the Transformer's $11\%$.
No significant deviations are observed between the training and the validation scores.
Two NRI blocks are consistently found to produce optimal results, with larger feed-forward dimensions, favouring $512$, and no additional block MLPs within.

\begin{figure}[ht]
    \centering
    \includegraphics[width=\columnwidth]{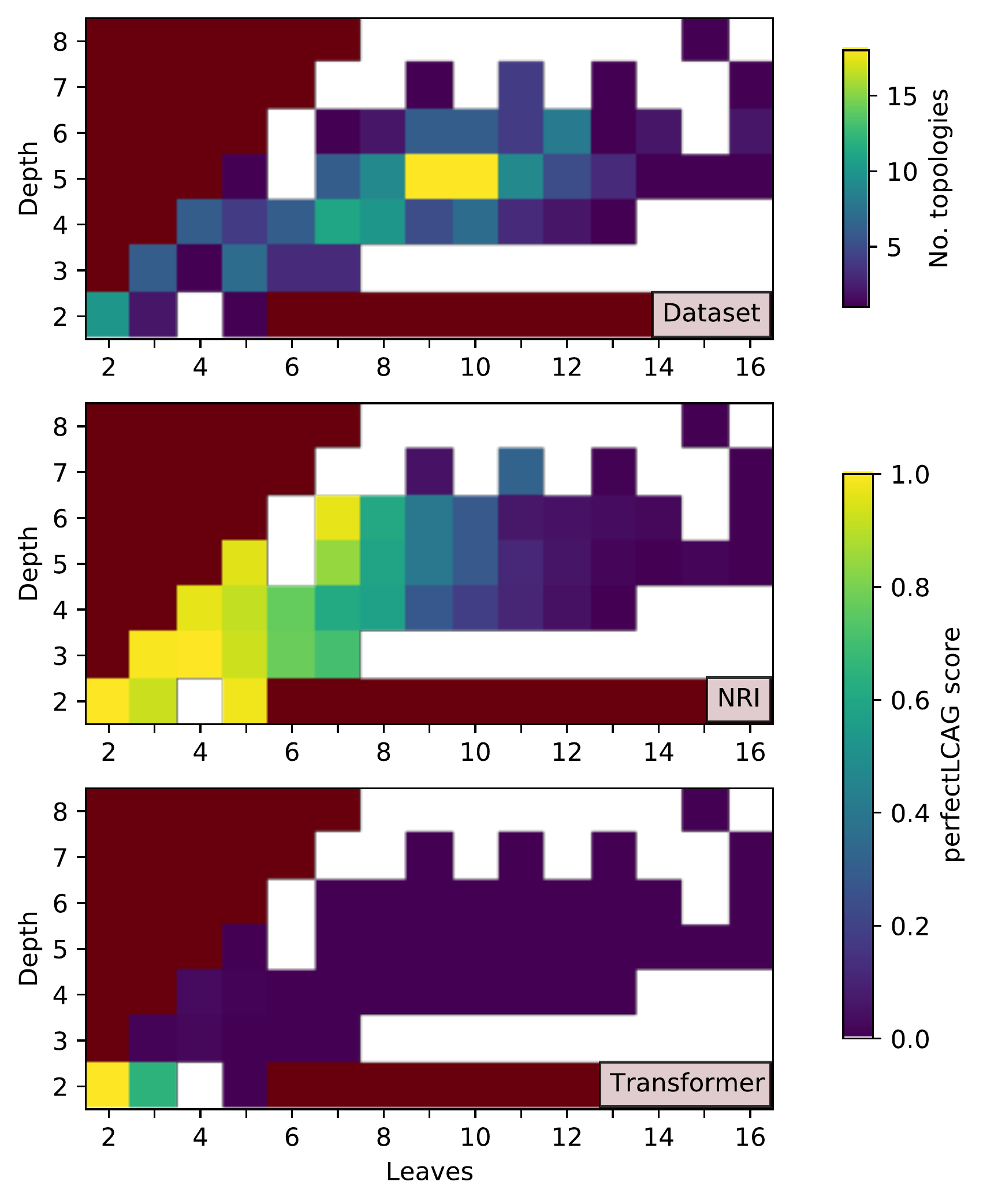}
    \caption{Dataset coverage (top) and $\mathrm{perfectLCAG}$ scores (middle, bottom) of the first experiment.
    Dark red indicates regions disallowed by the min/max children restrictions on topologies, white indicates regions with no training samples.
    Colours in the top plot show the number of topologies with the given depth/leaves configuration, colours in the lower two plots show the average $\mathrm{perfectLCAG}$ scores achieved on the given topologies' configurations.}
    \label{fig:phasespace-results-perfectLCAG}
\end{figure}

Based on these results, we performed a longer training for both the NRI and Transformer encoders using the optimal hyperparameters found and a random seed of $101$ until the validation $\mathrm{perfectLCAG}$ score no longer improved ($500$ and $323$ epochs, respectively).
The results on the test dataset are shown in \cref{fig:phasespace-results-perfectLCAG}, where the NRI and Transformer achieved an average $\mathrm{perfectLCAG}$ of $46.6\%$ and $5.80\%$, and a corresponding average accuracy of $91.6\%$ and $38.6\%$, respectively.
As with the hyperparameter optimisation, we observe no significant difference between the training and validation $\mathrm{perfectLCAG}$ scores, around $54\%$ and $10\%$ for the NRI and Transformer, respectively, but a clear sign of overfitting with respect to the test dataset.
Overall, the Transformer struggles to learn more than the most trivial topologies, whereas the NRI achieves reasonable $\mathrm{perfectLCAG}$ scores on most of those below $10$ leaves.
Looking closer, for trees up to and including $6$ leaves, the average $\mathrm{perfectLCAG}$ for the NRI is $92.5\%$, and for cases of trees up to and including $10$ leaves $59.7\%$.
The scores for all other subdivisions are shown in \cref{tab:test-results-leaves}.
This result demonstrates not only that the NRI is an appropriate choice of architecture for learning physical interactions for a variety of tree sizes, but also that it behaves as expected, namely by learning what we intuitively identify as simpler topologies first.

\begin{table*}[!htb]
    \caption{Average $\mathrm{perfectLCAG}$ scores for subsets of topologies of increasing complexity.}
    \label{tab:test-results-leaves}
    \centering
    \scriptsize
    \begin{tabularx}{\linewidth}{Xrrrrrrrrrrrrrrrr}
        \toprule
        \textbf{Model} & \multicolumn{15}{c}{\textbf{Average perfectLCAG (\%) score for leaves up to and including}} \\
         & 2 & 3 & 4 & 5 & 6 & 7 & 8 & 9 & 10 & 11 & 12 & 13 & 14 & 15 & 16 \\
        \midrule
        Transformer & \textbf{100.0} & 63.1 & 46.3 & 30.5 & 24.7 & 17.1 & 13.0 & 9.8 & 7.7 & 6.8 & 6.3 & 6.1 & 6.0 & 5.9 & 5.8 \\
        $\mathrm{NRI}$ & \textbf{100.0} & 98.8 & 98.2 & 96.2 & 92.5 & 85.7 & 79.2 & 68.5 & 59.7 & 54.3 & 50.3 & 48.8 & 48.1 & 47.6 & 46.6 \\
        $\mathrm{NRI}_{2048}$ & \textbf{100.0} & \textbf{99.3} & \textbf{98.7} & \textbf{97.5} & \textbf{94.2} & \textbf{87.1} & \textbf{79.7} & \textbf{70.1} & \textbf{60.9} & \textbf{55.7} & \textbf{51.5} & \textbf{50.0} & \textbf{49.2} & \textbf{48.7} & \textbf{47.7} \\
        $\mathrm{NRI}_{\mathrm{no blocks}}$ & \textbf{100.0} & 91.2 & 69.1 & 46.0 & 37.2 & 25.7 & 19.6 & 14.7 & 11.6 & 10.3 & 9.4 & 9.1 & 9.0 & 8.9 & 8.7 \\
        \bottomrule
    \end{tabularx}
\end{table*}

Interestingly, we observe no significant advantage to an increased number of topologies at a given complexity.
For example the large number of topologies at depth $5$ with leaves $9$ or $10$ in the dataset show no noticeable performance improvement over those at depth $6$ with the same number of leaves.
This indicates that the model is only able to learn each topology individually, and is unable to generalize to topologies of similar complexity.

\subsection{Ablation Studies}
\label{sec:phasespace-ablation}

Given the significantly better performance of the NRI over the Transformer, we explored the importance of the sequential edge-to-node and node-to-edge encoding layers within the NRI blocks.
To do so, we repeated the previous training with all blocks removed, leaving only the initial node-to-edge encoding layer to infer the LCAG from pair-wise combinations of the embeddings alone.
The results are shown in \cref{fig:NRI-ablation-no-blocks}, which reached an average test $\mathrm{perfectLCAG}$ of $8.7\%$, only slightly outperforming the Transformer.
We therefore conclude that the ability to iteratively combine node pairs is an essential feature of the NRI's ability to predict the LCAG, and similar features should be considered in any other Graph Neural Network approaches explored in future.

The hyperparameter optimisations demonstrated a clear preference for larger feed-forward dimensions and a lower number of initial/final feed-forward layers.
We therefore performed an additional hyperparameter search on these parameters up to a maximum feed-forward width of $2048$\footnote{Due to hardware constraints we were unable to explore larger widths.}, with all other hyperparameters set to the optimal found in the previous search.
The results are shown in \cref{fig:NRI-hyperparam-largedim} in the supplementary material, where no additional initial/final layers was found to be ideal, along with the maximum feed-forward width.

Following this finding, we repeated the training for $177$ epochs (until validation $\mathrm{perfectLCAG}$ no longer improved) with these hyperparameters and the same random seed as in the trainings in \cref{sec:phasespace-exp}.
The results of are shown in \cref{fig:NRI-ablation-2048}.
We observe an average $\mathrm{perfectLCAG}$ of $47.7\%$ ($94.2\%$ and $60.9\%$ for the $\leq 6$ and $\leq 10$ regions, respectively) and accuracy of $92.6\%$, which represents only a small improvement over the previous results.
Detailed results are shown in \cref{tab:test-results-leaves}.
We see a relatively large discrepancy between this score and the maximum hyperparameter optimisation $\mathrm{perfectLCAG}$ reached in \cref{fig:NRI-hyperparam-largedim} in the supplementary material of $59.6\%$ on the validation dataset.
Similar to the experiments in \cref{sec:phasespace-exp}, we also find a comparable discrepancy in the validation $\mathrm{perfectLCAG}$ score of the $177$ epoch training, which achieved $58.4\%$.

To summarise, with a width of 1024, the model achieved $46.6\%$ test and $46\%$ validation $\mathrm{perfectLCAG}$, and for 2048 with all else equal it reached $47.7\%$ test and $59.6\%$ validation $\mathrm{perfectLCAG}$. 
We therefore conclude that the hyperparameter optimisation has caused overfitting to the validation dataset, and that an increase in the fully-connected widths within the NRI only serves to exacerbate this problem.

\subsection{Discussion}
\label{sec:discussion}

While the Belle~II experiment typically has ten or fewer leaves, the noisy experimental environments in hadron colliders like the LHC mean that many other particles can be produced, raising the total number of leaves significantly.
However, it is often the case that only a subset of leaves that belong to the particle decay of interest are selected.
For example, the LHCb triggers~\cite{aaijLHCbTriggerIts2013,gligorovEfficientReliableFast2013} have a specific focus on filtering for events with $b$-hadrons whose decay products are subsequently analysed.
As our results demonstrate high performance for scenarios with up to ten leaves, our proposed approach already offers utility in existing experimental scenarios.
Nevertheless, there is still significant room for improvement, the key to which we believe will be finding exactly which hyperparameters can be modified to improve performance for more leaves.

\begin{figure}[ht]
\vskip 0.2in
\begin{center}
\begin{subfigure}[b]{\columnwidth}
\centerline{
\includegraphics[width=0.95\columnwidth]{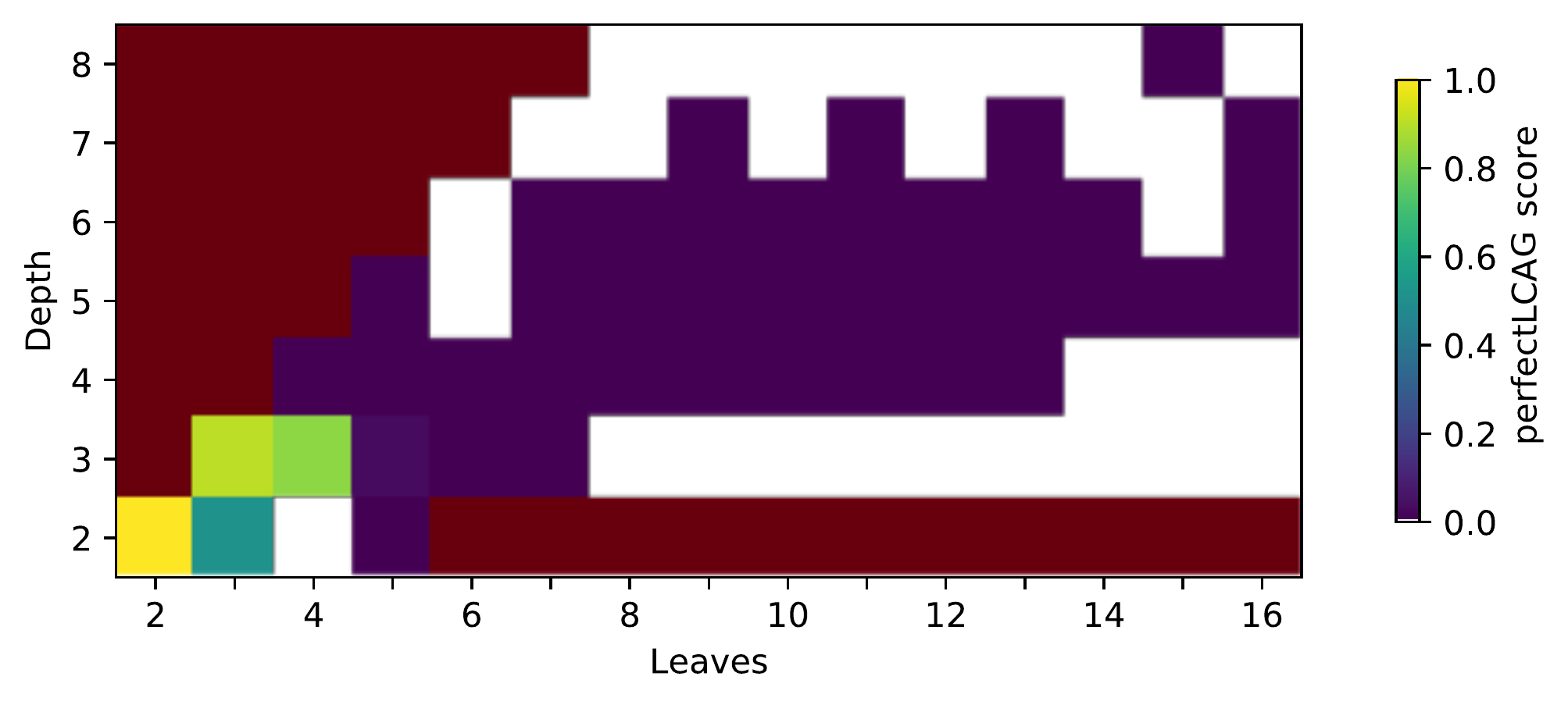}
}
\caption{
    NRI encoder without NRI blocks (trained using a single \texttt{Node to Edge}).
}
\label{fig:NRI-ablation-no-blocks}
\end{subfigure}
\begin{subfigure}[b]{\columnwidth}
\centerline{
\includegraphics[width=0.95\columnwidth]{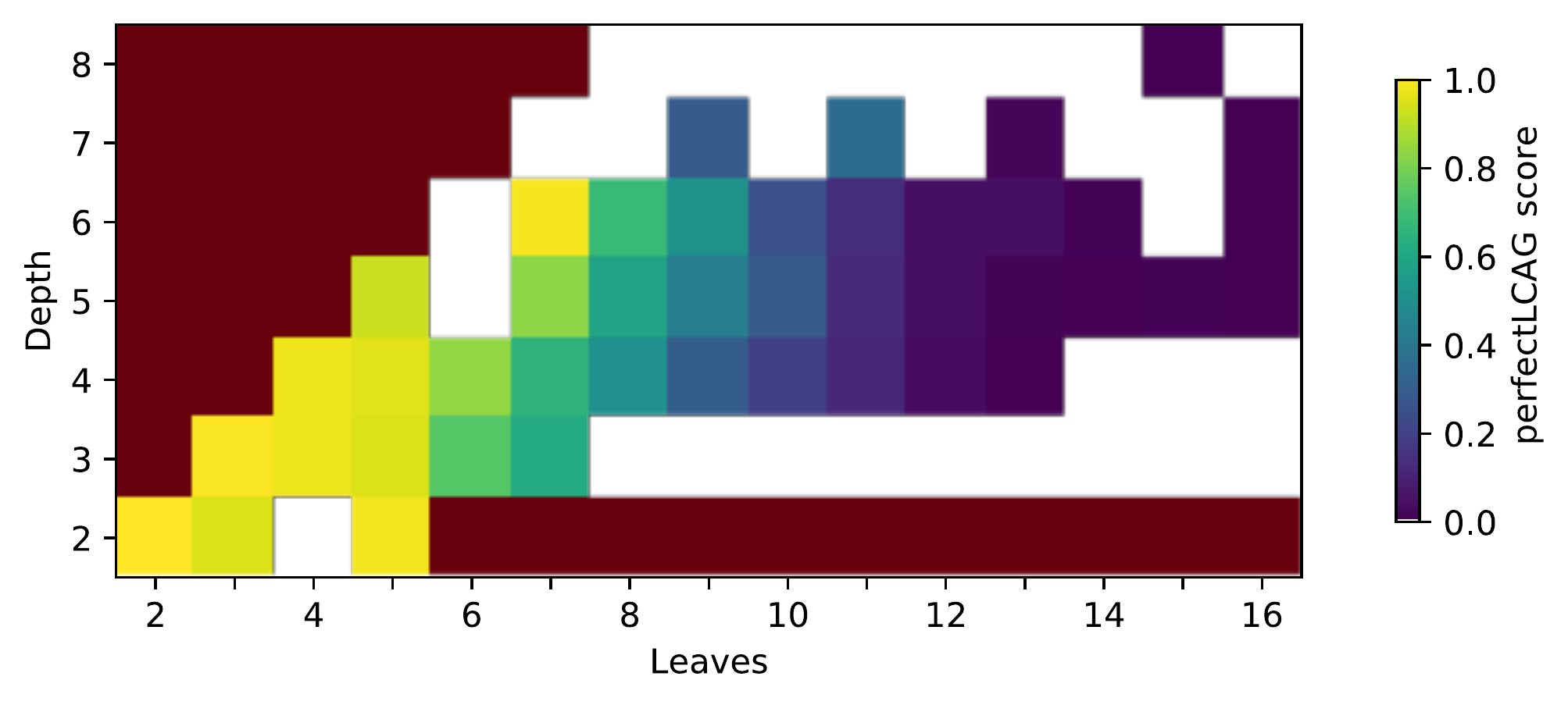}
}
\caption{
    NRI encoder with a feed-forward layer width of $2048$ for all MLP components.
}
\label{fig:NRI-ablation-2048}
\end{subfigure}
\caption{
    Ablation study results for the NRI encoder on the test dataset shown in \cref{fig:phasespace-results-perfectLCAG} (top). 
    The network failed to learn any complex topologies.
}
\label{fig:NRI-ablation}
\end{center}
\end{figure}

We have already explored some of the hyperparameters in the experiments and shown the essential role of the NRI blocks (\cref{fig:NRI-ablation}).
We also demonstrated that a larger model capacity (larger width of feed-forward layers) corresponds to improved performance (\cref{tab:test-results-leaves}), but saw that this quickly leads to overfitting.
While a further investigation into this overfitting is outside the scope of this work, as our primary interest was to perform an initial exploration into what appear to be the essential NRI hyperparameters, based on these findings we recommend considering the inclusion of cross-validation, stronger regularization, or ensemble methods to identify and mitigate this.

Looking further, we saw that the choice of architecture has the largest impact on performance, with the Transformer appearing unable to learn a meaningful solution (\cref{fig:phasespace-results-perfectLCAG}).
We concluded in \cref{sec:phasespace-ablation} that the NRI's ability to iteratively combine node pairs was an essential component for its performance, however graph nodes do not exist only as child pairs, but also triplets and more.
Therefore, a broad investigation into various other architectures, especially Graph Neural Networks, with potentially better inductive biases is likely the most promising avenue for increasing performance to handle more leaves and larger trees.

\section{Conclusion}
\label{sec:conclusion}

Learning tree structures from leaves is a highly relevant, yet largely unexplored problem from a machine learning point of view, particularly for physical processes such as those in the field of particle physics.
Recent years have seen not only the rise of machine learning in modeling the dynamics of physical systems, but more specifically the use of Graph Neural Networks (GNNs) to do so.
This shift represents an understanding that the representation of input data, as unordered sets with relational information, is key to enabling such learning.
Despite this, there has been little focus on constructing equally meaningfully representations of the prediction targets.

In this work, we have introduced the lowest common ancestor generation (LCAG) matrix as a novel, compact representation of tree-structured data.
We have shown that the LCAG, while equivalent to the adjacency matrix, resolves the constraints that make learning the adjacency matrix directly impractical.
By doing so, the LCAG enables the first end-to-end trainable solution which learns the hierarchical structure of varying tree sizes directly, using only the terminal tree leaves to do so.

We have demonstrated the use of the LCAG in the task of learning to predict a variety of particle decay processes on a simulated physics dataset.
Our results show that when selecting an appropriate GNN, in this case the Neural Relational Inference (NRI) encoder, the network is able to correctly predict the LCAG for $92.5\%$ of decay trees up to and in including $6$ leaves, and $59.7\%$ of trees up to and including $10$ leaves.
We showed how the NRI blocks are an essential contributor to these results, and that the Transformer was unable to converge on a meaningful solution.
Our results hint at the possibility of other GNN with potentially better inductive biases enabling the performance to scale to larger trees with more leaves.

Our results pave the way to construct end-to-end trainable neural approaches to particle decay reconstruction at large collaborations such as the LHC or Belle~II that may eventually replace existing algorithms.
While existing particle physics reconstruction algorithms could in theory be adapted to the dataset used in this work for a direct comparison, this would entail very significant work as they have been heavily tailored towards the context of their respective experiments.
It is both simpler and more meaningful for the experiments to instead explore the LCAG approach on real-world physics datasets directly.

While this work has only scratched the surface of what can be achieved when predicting the LCAG, particularly when exploring other graph interaction learning neural network architectures, we believe it to be a powerful representation tool as it resolves many of the constraints on existing hierarchical learning approaches.
We expect that this work will serve as inspiration not only to high-energy particle physics experiments, but other domains working with similar problems and tree-structured data as well.

\section*{Acknowledgments}

This work is supported by the Helmholtz Association Initiative and Networking Fund under the Helmholtz AI platform grant and the HAICORE@KIT partition, the Bundesministerium f\"{u}r Bildung und Forschung (BMBF) under the grant 05H21PDKBA, and the  L’Institut National de Physique Nucléaire et de Physique des Particules (IN2P3) du CNRS (France) and the French Agence Nationale de la Recherche (ANR) under grant ANR-21-CE31-0009 (project FIDDLE) and the Seed Money programme of Eucor – The European Campus.
We wish to also thank Julián García Pardiñas and Isabelle Ripp-Baudot for the fruitful discussions.

\printbibliography
\clearpage

\appendix
\onecolumn

\section*{Supplementary Material}

\subsection*{Hyperparameter Searches}

\begin{figure*}[b]
    \centering
    \includegraphics[width=1.0\linewidth]{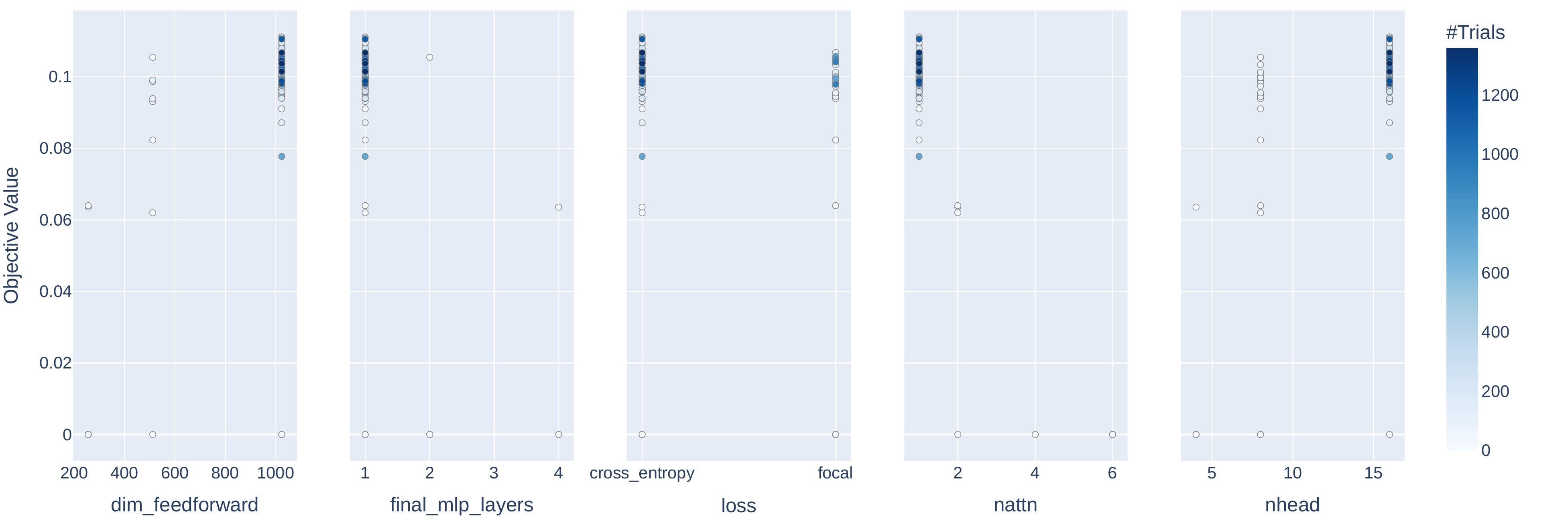}
    \caption{Hyperparamter optimisation results for the Transformer encoder model.
    The objective value optimized for is the validation $\mathrm{perfectLCAG}$ score.}
    \label{fig:transformer-hyperparam}
\end{figure*}

\begin{figure*}[b]
    \centering
    \includegraphics[width=1.0\linewidth]{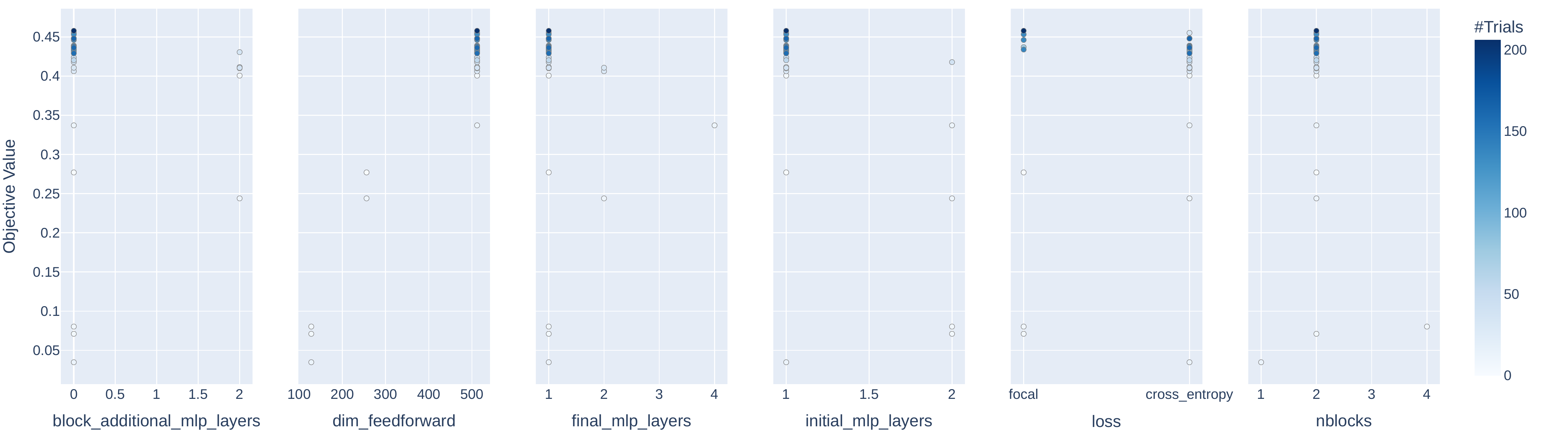}
    \caption{Hyperparamter optimisation results for the NRI encoder model.
    The objective value optimized for is the validation $\mathrm{perfectLCAG}$ score.}
    \label{fig:NRI-hyperparam}
\end{figure*}

\begin{figure*}[b]
    \centering
    \includegraphics[width=0.5\linewidth]{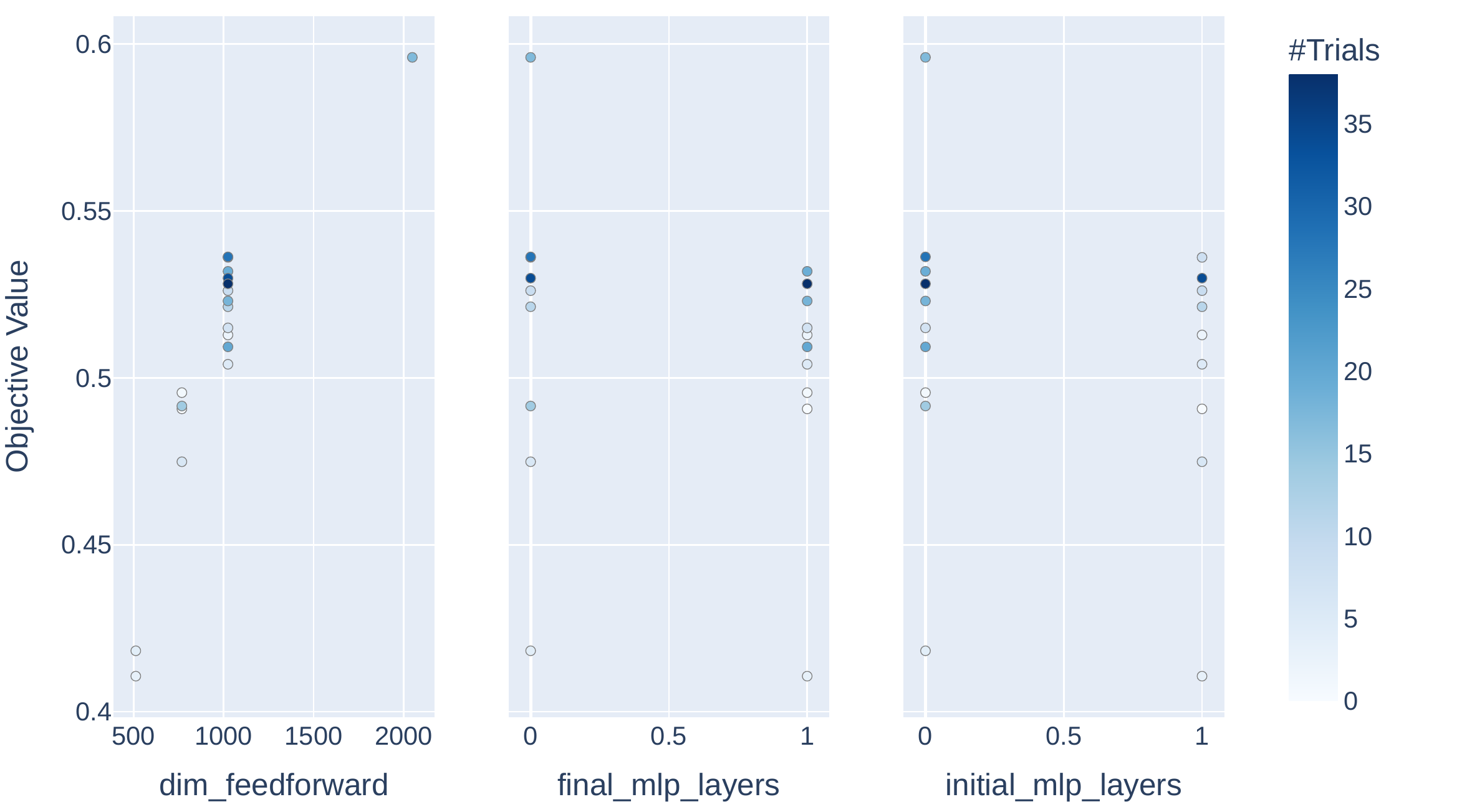}
    \caption{Hyperparamter optimisation results when tuning for larger NRI feed-forward layer widths.
    Note the clear trend of increasing objective value (validation $\mathrm{perfectLCAG}$ score) with increasing layer width within blocks (\texttt{dim\_feed-forward}).
    The objective value optimized for is the validation $\mathrm{perfectLCAG}$ score.}
    \label{fig:NRI-hyperparam-largedim}
\end{figure*}

\end{document}